\documentclass[]{aastex63}

\usepackage{array,multirow}
\usepackage{amsmath}
\usepackage{color,soul}
\usepackage{textcomp}
\usepackage{booktabs}
\usepackage{CJK}

\received{May 7, 2021}
\revised{Oct 2, 2021}
\accepted{Nov 3, 2021}
\submitjournal{PSJ}

\shorttitle{A Comparison Study of Tholins: Surface Energy}

\graphicspath{{./}{figures/}}

\begin{document}

\title{A Cross-Laboratory Comparison Study of Titan Haze Analogs: Surface Energy}

\correspondingauthor{Xinting Yu}
\email{xintingyu@ucsc.edu}

\author[0000-0002-8110-7226]{Jialin Li\begin{CJK*}{UTF8}{gbsn}
(李嘉霖)\end{CJK*}}
\affiliation{Department of Physics, University of California Santa Cruz\\
1156 High St\\
Santa Cruz, CA 95064, USA}

\author[0000-0002-7479-1437]{Xinting Yu\begin{CJK*}{UTF8}{gbsn}
(余馨婷)\end{CJK*}}
\affiliation{Department of Earth and Planetary Sciences, University of California Santa Cruz\\
1156 High St\\
Santa Cruz, CA 95064, USA}

\author[0000-0002-1883-552X]{Ella Sciamma-O'Brien}
\affiliation{NASA Ames Research Center, Space Science \& Astrobiology Division, Astrophysics Branch\\
Moffett Field, CA 94035, USA}

\author[0000-0002-6694-0965]{Chao He}
\affiliation{Department of Earth and Planetary Sciences, Johns Hopkins University\\
3400 N. Charles Street\\
Baltimore, MD 21218, USA}

\author[0000-0001-9612-8532	]{Joshua A. Sebree}
\affiliation{Department of Chemistry and Biochemistry, University of Northern Iowa\\ 
1227 W 27th St\\
Cedar Falls, IA 50614, USA}

\author[0000-0002-6064-4401]{Farid Salama}
\affiliation{NASA Ames Research Center, Space Science \& Astrobiology Division, Astrophysics Branch\\
Moffett Field, CA 94035, USA}

\author[0000-0003-4596-0702]{Sarah M. H\"{o}rst}
\affiliation{Department of Earth and Planetary Sciences, Johns Hopkins University\\
3400 N. Charles Street\\
Baltimore, MD 21218, USA}

\author{Xi Zhang}
\affiliation{Department of Earth and Planetary Sciences, University of California Santa Cruz\\
1156 High St\\
Santa Cruz, CA 95064, USA}

\begin{abstract}
In Titan's nitrogen-methane atmosphere, photochemistry leads to the production of complex organic particles, forming Titan’s thick haze layers. Laboratory-produced aerosol analogs, or ``tholins", are produced in a number of laboratories; however, most previous studies have investigated analogs produced by only one laboratory rather than a systematic, comparative analysis. In this study, we performed a comparative study of an important material property, the surface energy, of seven tholin samples produced in three independent laboratories under a broad range of experimental conditions, and explored their commonalities and differences. All seven tholin samples are found to have high surface energies, and are therefore highly cohesive. Thus, if the surface sediments on Titan are similar to tholins, future missions such as Dragonfly will likely encounter sticky sediments. We also identified a commonality between all the tholin samples: a high dispersive (non-polar) surface energy component of at least 30 mJ/m$^2$. This common property could be shared by the actual haze particles on Titan as well. Given that the most abundant species interacting with the haze on Titan (methane, ethane, and nitrogen) are non-polar in nature, the dispersive surface energy component of the haze particles could be a determinant factor in condensate-haze and haze-lake liquids interactions on Titan. With this common trait of tholin samples, we confirmed the findings of a previous study by \citet {2020ApJ...905...88Y} that haze particles are likely good cloud condensation nuclei (CCN) for methane and ethane clouds and would likely be completely wetted by the hydrocarbon lakes on Titan. \\\\
\end{abstract}


\section{Introduction} \label{sec:intro}
Titan, the largest moon of Saturn, is known for its thick and hazy atmosphere filled with rich and complex organic materials. Thanks to the Cassini-Huygens mission, the most recent mission to perform in-situ measurements on Titan, surface features such as the organic dunes in the equatorial regions (e.g., \citealt*{2006Sci...312..724L}) and the presence of liquid hydrocarbon lakes in the polar regions (e.g., \citealt*{2007Natur.445...61S}) were discovered. The Cassini-Huygens spacecraft also characterized various properties of Titan's atmosphere (see \citealt*{2017JGRE..122..432H} for a review, and references therein). Titan's atmosphere is composed of predominantly N$_2$, with 1--5\% of CH$_4$, small amounts of CO, as well as trace amounts of various hydrocarbons and nitriles \citep{2005Sci...308..975F,2005Natur.438..779N}. A number of energy sources including energetic particles and solar UV photons initiate chemical reactions in Titan's upper atmosphere, resulting in the formation of complex organic aerosol particles that form the thick haze layers (\citealt{2017JGRE..122..432H}, and references therein). These haze particles can act as cloud condensation nuclei (CCN) for various condensable simple organic molecules to form clouds \citep{2006Icar..182..230B,2008Icar..195..792C,2011Icar..215..732L,2019A&A...631A.151R}. They can also fall to the surface where they partake in fluvial and aeolian processes. The majority of the dune materials and the sediments in the polar lakes are believed to be derived from the aerosol particles formed in the atmosphere \citep{2007P&SS...55.2025S,2020NatAs...4..228L}. 

One of the main efforts to better understand the chemistry of Titan's haze is through making synthetic analogs of Titan's aerosols, so-called ``tholins", in a laboratory setting. Since the commencement of such experiments 40 years ago \citep{1979Natur.277..102S}, several groups have developed laboratory setups to produce tholins and have analyzed their various physical, chemical, and optical properties (see reviews by \citealt{2012ChRv..112..1882,2013P&SS...77...91C,2017JGRE..122..432H}, and references therein; as well as \citealt{2017Icar..289..214S,2017ApJ...841L..31H}, and \citealt{2018ApJ...865..133S}). In these experiments, the production of tholins usually involves the dominant gas constituents in Titan's atmosphere, N$_2$ and CH$_4$, and an energy source to trigger the chemistry, most often UV or charged particle irradiation. UV lamps (e.g., \citealt{2000Icar..147..282C,2006PNAS..10318035T,2018ApJ...865..133S}) and synchrotron sources (e.g., \citealt{2007GeoRL..34.2204I, 2009JPCA..11311211T,2013JGRE..118..778P}) can simulate the solar UV radiation of the mid- to upper atmosphere, while gamma rays/soft x-rays mimic protons from Saturn's magnetosphere and cosmic rays in the lower atmosphere (e.g., \citealt{2001AdSpR..27..261R,2012MNRAS.423.2209P}). Plasma discharges can simulate the electron bombardment of Titan's upper atmosphere from Saturn's magnetosphere and have a broad energy spectrum resembling the solar spectrum, with a high energy tail that allows for the dissociation of nitrogen (e.g., \citealt{1999P&SS...47.1331C,2006P&SS...54..394S,2012JPCA..116.4760H,2017Icar..289..214S,2017ApJ...841L..31H}). The choice of energy source as well as other experimental parameters like the molecular mixing ratio, pressure, or temperature used in the laboratory can have an impact on chemical and material properties of the tholin samples (e.g., \citealt{2010Icar..209..704S,2014Icar..243..325S, 2004Icar..168..344I,2018Icar..301..136H, 2012Icar..221..670M, 2014PP&P..11..409}). 

A prevailing question of the field is whether the laboratory simulated tholins can actually represent the actual haze particles on Titan. A number of laboratory-produced tholin samples underwent various characterization of their physical and chemical properties. Previous studies have found that laboratory-produced tholins are considered decent analogs to the haze particles in Titan's atmosphere as their chemical and optical properties match, at least partially, the Huygens Aerosol Collector Pyrolyzer (ACP) measurements \citep{2005Natur.438..796I}, the Huygens Descent Imager/Spectral Radiometer (DISR) measurements \citep{2008P&SS...56..669T,2010Icar..210..832L}, and the Cassini Visible and Infrared Mapping Spectrometer (VIMS) observations \citep{2012Icar..221..320G,2017Icar..289..214S}. The analysis of the gaseous molecular precursors that tholins are formed from is also in agreement with the Cassini Plasma Spectrometer (CAPS) measurements \citep{2014Icar..243..325S,2019ApJ...872L..31D} and the Cassini Ion and Neutral Mass Spectrometer (INMS) measurements \citep{2014Icar..243..325S, 2020Icar..33813437D}. However, these studies are based on tholin samples produced in specific laboratories. Only a few studies have been done to compare tholin samples made in different laboratories. \citet{2012ChRv..112..1882} reviewed extensively all aspects of the tholins reported in literature during the Cassini-Huygens mission, including the composition, various chemical and physical properties, and production methods. They were able to derive a metric for determining the most Titan-like analogs based on four parameters: energy source, temperature, pressure, and energy density, and showed that different experimental conditions and experimental setups can lead to tholins with different properties. \citet{2013P&SS...77...91C} measured volatiles produced after a thermal treatment for a few tholin samples produced with different energy sources in different laboratory setups and compared the results to the data of the Huygens ACP. This study demonstrated that the properties of one specific tholin sample cannot be assumed to be representative of the actual Titan haze as they might be dependent on the experimental conditions used to produce the sample. \citet{2014E&PSL.403...99C} conducted a comparative analysis of tholin samples produced from three different experimental set-ups and identified the amines common to all three samples. \citet{2015P&SS..109..159B} reviewed and summarized the measured optical properties of tholins produced by various laboratories with different experimental setups.

The few comparative studies listed above were heavily focused on the chemical and optical properties of tholins, and only a few studies have investigated the commonalities/differences between tholins and actual aerosols on Titan \citep{2010Icar..210..832L,2013P&SS...77...91C,2015P&SS..109..159B}. No systematic comparative studies have been conducted however to investigate the material properties of tholin samples produced by different laboratories with different experimental conditions, even though material properties play important roles in governing various atmospheric and surface processes on Titan \citep{2017JGRE..122.2610Y,2018JGRE..123.2310Y,2020ApJ...905...88Y,2020E&PSL.53015996Y}.

In this study, we characterized an important bulk physiochemical property, the surface energy, of tholin samples made by different laboratory groups with distinct laboratory setups and experimental conditions. The goal of this investigation was to see if common traits could be found between different laboratory-made aerosol analogs. If there were traits shared by all tholins, then it could be expected that Titan aerosols present such traits as well. The surface energy is defined as the change of free energy when the surface area of a solid is increased a unit area. It has a unit of J/m$^2$. This free energy change is equivalent to the energy needed to separate two contacting solid surfaces per two unit area and it determines the adhesion and wettability of the surfaces \citep{2011Israelachvili}. The surface energy can be extracted through a simple measurement technique, the sessile drop contact angle method, and it can reveal the fundamental bulk chemical make-up, cohesiveness, and wetting properties of a material. The surface energy of Titan haze particles has important implications for cloud formation, aerosol-lake interactions, sand transport and dune formation on Titan's surface \citep{2020ApJ...905...88Y}. Note that the equivalent physiochemical property for the liquid is referred to as the ``surface tension", which has a unit of N/m (equilvalent to J/m$^2$). Throughout the text, we use ``surface free energy" when we refer to the surface energy of the solid and the surface tension of the liquid at the same time.

In the study presented here, we measured the surface energies of seven tholin samples produced in three laboratory facilities simulating planetary atmospheric chemistry, and performed an in-depth comparison study between these samples. The three laboratories are respectively the Photochemical Aerosol Chamber (PAC) facility at the University of Northern Iowa, or UNI (e.g., \citealt{2018ApJ...865..133S}), the Planetary Haze Research facility (PHAZER) at the Johns Hopkins University, or JHU (e.g., \citealt{2017ApJ...841L..31H}), and the COSmIC facility at the NASA Ames Research Center, or ARC (e.g., \citealt{2017Icar..289..214S}). Below we describe the three laboratory setups and the experimental protocols used to produce the tholin samples in these facilities, the contact angle measurements, and the surface energy derivation method, before summarizing the surface energy results and comparing the similarities and the differences between the seven tholin samples studied. The common traits of tholins and their implications for processes occurring on Titan are then discussed. 

\section{Experimental} \label{sec:exper}

\subsection{Production of Tholin Samples} \label{subsec:materials}

\begin{figure}[ht] 
    \centering
    \includegraphics[width=\textwidth]{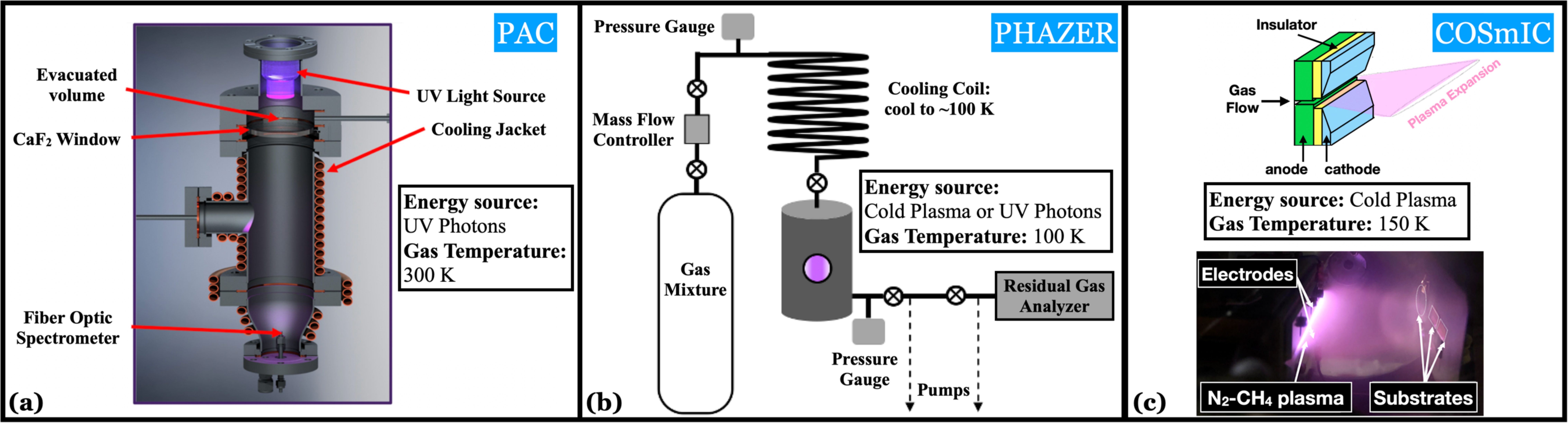}
    \caption{Schematic of the three experimental setups used to produce tholins in this study: (a) the PAC facility at UNI, (b) the PHAZER facility at JHU, and (c) the COSmIC facility at NASA ARC.}
    \label{fig:facility}
\end{figure}

In this study, we characterized seven tholin samples produced by three laboratories with different experimental setups and conditions: the PAC chamber at UNI, the PHAZER chamber at JHU, and the COSmIC chamber at ARC (Figure \ref{fig:facility}). Doing comparative studies of samples produced in different experimental facilities is not trivial as each facility will have their own characteristic and many experimental parameters can be different from one setup to the other. A systematic analysis allowing to investigate the effects of a limited number of experimental parameters takes time and becomes more and more difficult when increasing the number of facilities to be compared. In the study presented here, we considered laboratory setups that use two different types of energy sources to simulate Titan's atmospheric chemistry: cold plasma discharges and UV/deuterium lamps. The UV lamps simulate the solar UV irradiation above 115~nm and the cold plasma discharges simulate energetic particle bombardment from Saturn's magnetosphere. These processes are the two main drivers of chemistry in Titan's atmosphere. By doing a comparative study of tholin samples produced in these three laboratories we can compare tholins produced from plasma chemistry in two different setups (JHU/PHAZER and ARC/COSmIC) and tholins produced from UV chemistry in two different setups (JHU/PHAZER and UNI/PAC) as detailed below.

To investigate the effect of the energy source on the surface properties of tholins produced in different experimental setups, four tholin samples (two from UV chemistry and two from plasma chemistry) were produced from the same N$_2$/CH$_4$ (95/5) gas mixture in the three laboratories. Because the JHU/PHAZER setup allows the use of both energy sources to produce tholins, we could also investigate the effect of the energy source on the surface properties of two tholin samples produced within a specific experimental setup. Similarly, the COSmIC setup was also used to produce three additional tholin samples using different gas mixtures, in order to investigate the effects of initial gas mixtures on the surface properties of tholin samples produced with the same laboratory setup. Details on the seven tholin samples characterized in this study are provided in Table 1. From now on, the samples will be referred to by the sample names listed in this table. The study presented here is one of only a few studies that has compared tholins from different laboratories and we are planning to conduct more experiments with these laboratories to examine the effect of gas mixtures and other parameters in-depth. The encouraging results presented here will hopefully lead to more comparative studies involving more experimental setups in the future.

\begin{table}
\caption{List of the seven different tholin samples used in this study, the details of the experimental setups they were produced in, and the associated production conditions.}
\label{table:tholin_samples}
\centering
\hspace*{-4cm}
\begin{tabular}{crcrcrcrcrcrcrcrc}
\toprule
Laboratory && Sample &&  Energy&& Reaction && Gas && Gas  && Gas && Pressure&& Gas flow \\

location && name &&  source && time && mixing  && exposure && temp && (Pa) && rate\\

(Setup) &&&&&& (hr) &&ratio&&time&&(K)&&&&(sccm)\\
\cmidrule{1-17}

UNI &&\multirow{2}{*}{UNI-UV} && Deuterium lamp &&\multirow{2}{*}{72}&& N$_2$:CH$_4$&&\multirow{2}{*}{3 min}&&\multirow{2}{*}{300}&&\multirow{2}{*}{6.7} $\times$10$^{4}$ &&\multirow{2}{*}{20}\\
(PAC) &&&& 115--400 nm&&&&(95:5)&&&&&&&&\\

\cmidrule{1-17}

&& \multirow{2}{*}{JHU-UV} && Hydrogen UV lamp &&\multirow{2}{*}{144} && && \multirow{4}{*}{3 s}&& \multirow{4}{*}{100}&& \multirow{4}{*}{267} &&\multirow{4}{*}{10}\\
JHU & &&& 115--400 nm&& && N$_2$:CH$_4$&&&&&&&&\\
 \cmidrule{3-3}\cmidrule{5-5}\cmidrule{7-7}
(PHAZER) && \multirow{2}{*}{JHU-Plasma}&&  AC plasma && \multirow{2}{*}{72}&& (95:5) &&&&&&\\
&&&& discharge &&&&&&&&&&&&\\

\cmidrule{1-17}

&&\multirow{2}{*}{ARC-Plasma-N$_2$-1}&&&& \multirow{2}{*}{13} && N$_2$:CH$_4$ &&\multirow{8}{*}{3.5\ $\mu$s}&&\multirow{8}{*}{150}&&\multirow{8}{*}{3000} &&\multirow{8}{*}{2000}\\
&& &&&& && (95:5)&&&&&&&&\\

\cmidrule{3-3}\cmidrule{7-7}\cmidrule{9-9}
NASA && \multirow{2}{*}{ARC-Plasma-N$_2$-2}&&&& \multirow{2}{*}{5} && N$_2$:CH$_4$:C$_2$H$_2$&&&&&&&&\\

ARC &&&&DC plasma&&&&(94.5:5:0.5)&&&&&&&&\\
\cmidrule{3-3}\cmidrule{7-7}\cmidrule{9-9}
(COSmIC) && \multirow{2}{*}{ARC-Plasma-Ar-1}&&  discharge&& \multirow{2}{*}{9.5} && Ar:CH$_4$&&&&&&&&\\
&& &&  && && (95:5)&&&&&&&&\\

\cmidrule{3-3}\cmidrule{7-7}\cmidrule{9-9}
&& \multirow{2}{*}{ARC-Plasma-Ar-2}&&&& \multirow{2}{*}{5} && Ar:CH$_4$:C$_2$H$_2$ &&&&&&&&\\

&&&&&&&&(94.9:5:0.1)&&&&&&&&\\
\bottomrule
\end{tabular}
\end{table}

\subsubsection{Photochemical Aerosol Chamber (PAC) at UNI}
The PAC facility at UNI was designed and built to simulate the photochemistry occurring in Titan's upper atmosphere using UV photons as the energy source (schematic shown in Figure \ref{fig:facility}a). The PAC reaction chamber allows to run experiments over a range of gas temperatures (200--300 K), pressures (0.1--1000 Torr --- 10--10$^5$ Pa), and gas phase reaction precursors. High-pressure gas mixtures ($\sim$5.5$\times$10$^{6}$ Pa) are made in a 2 L stainless steel gas mixing manifold, which enables the study of reactants at concentrations ranging from 10 ppb--10\% in N$_2$. After the gas mixture homogenizes, it can flow continuously at a flow rate of up to 2000 standard cubic centimeters per minute (sccm) directly into the UV-reaction chamber. The UV energy source is a Lyman-alpha deuterium lamp (Hamamatsu L11798) with an MgF$_2$ window, which produces a continuum of UV photons from 115 to 400 nm, with a UV flux of $\sim$5.0$\times$10$^{15}$ photons/s \citep{2018ApJ...865..133S}. The UV photons then pass through the PAC chamber's CaF$_2$ window before irradiating the gases. The gas mixture continuously flows through the reaction chamber and chemistry is induced by UV irradiation of the gas mixture, resulting in the formation of aerosols. The formed aerosols are carried by the flow of gas from the reaction chamber down to a collection chamber where tholins are collected on a substrate.

To produce the UNI-UV tholin sample used in the study presented here, ultra high purity N$_2$ (99.95\%) and CH$_4$ (99.995\%) gases were mixed at a mixing ratio of 95:5 in the gas manifold. The gas mixture was then flown into the reaction chamber at room temperature (300 K), with a pressure of 500 Torr (6.7$\times$10$^{4}$ Pa) and a flow rate of 20 sccm, which resulted in the gas mixture being exposed to UV irradiation for 3 minutes while it flowed through the reaction chamber. The tholins produced were deposited on a silicon window (Thorlabs WG80530) in the collection chamber. To produce this UNI tholin sample, the experiment was run for $\sim$72 hours.

\subsubsection{Planetary HAZE Research Chamber (PHAZER) at JHU}
The PHAZER facility at JHU was designed and built to simulate haze formation in planetary and exoplanetary atmospheres (schematic shown in Figure \ref{fig:facility}b). The system is a stainless steel chamber with copper gasket seals and can achieve an ultra-high vacuum (base pressure 10$^{-4}$--10$^{-5}$ Torr --- $\sim$10$^{-2}$--10$^{-3}$ Pa). PHAZER is equipped with two energy sources: a cold plasma generated by an AC glow discharge \citep{2017ApJ...841L..31H} and a hydrogen UV lamp \citep{2018ApJ...856L...3H} with an MgF$_2$ window, which creates a continuum of UV photons from 115 to 400 nm (HHeLM-L, Resonance LTD) with a total UV flux of 1.4$\times$10$^{15}$ photons/s. The system can be operated at a broad range of temperatures (100--800 K) as well as pressures (0.1--10 Torr --- $\sim$10--10$^{3}$ Pa) for the cold plasma source and 0.1-1000 Torr --- $\sim$10--10$^{5}$~Pa) for the UV source), and with a variety of reactant gases. The PHAZER facility has been successfully used for a series of planetary and exoplanetary atmospheric simulation experiments, including for Titan and Pluto \citep{2017ApJ...841L..31H}, and for sub-Neptune exoplanets \citep{2018ApJ...856L...3H,2018AJ....156...38H,2020NatAs...4..986H,2020PSJ.....1...51H, 2018Icar..301..136H}.

To produce the JHU-UV and JHU tholin samples used in this study, pure N$_2$ (99.9997\%) and CH$_4$ (99.999\%) were flown through two mass flow controllers (MFC -- MKS Instrument, GM50A) to generate a N$_2$:CH$_4$ gas mixture with a mixing ratio of 95:5, flowing at 10 sccm. The gas mixture was first flown continuously through a 15-meter stainless steel coil immersed in a liquid nitrogen cold bath (77 K), in order to cool down the gases to $\sim$100 K and remove impurities \citep{2017ApJ...841L..31H}. The cooled gas mixture was then flown into the reaction chamber, generating a constant pressure of 2 Torr (267 Pa) in the chamber. While flowing through the reaction chamber, the gas mixture was exposed to one of the two energy sources, resulting in the production of tholins. The exposure time of the gas mixture to the energy source before being pumped out of the reaction chamber was 3 s. Acid-washed glass slides placed on the bottom flange inside the reaction chamber were used as substrates to collect the tholin samples. After running the experiment for 72 hours for the plasma experiment and 144 hours for the UV irradiation experiment, the substrates were coated with the tholin sample. These long running times ensured a homogeneous coating with sufficient thickness ($>$ 50 nm, \citealt{2020ApJ...905...88Y}) for the contact angle measurements to be independent of the coating thickness (e.g., \citealt{2005JAdh....81...29}).

\subsubsection{Cosmic Simulation Chamber (COSmIC)}
The COSmIC facility at NASA ARC was developed to simulate interstellar, circumstellar, and planetary environments at low temperature in the laboratory. It has been successfully used to produce and study analogs of gas phase neutral molecules and ions, and solid particles forming in interstellar and circumstellar environments \citep{2008IAUS..251..357S, 2013ApJS..208....6C,2020ApJ...889..101G,2020ApJ...905...45S}, as well as in Titan's atmosphere \citep{2014Icar..243..325S,2015IAUFM..29A.327S,2017Icar..289..214S,2018ApJ...853..107R}. The COSmIC facility uses a pulsed discharge nozzle (PDN) \citep{2006FaDi..133..289B,2005PhRvE...71..036409}; see schematic and picture in Figure \ref{fig:facility}c) coupled to a vacuum chamber (base pressure: 8 mTorr --- 1 Pa) to generate a pulsed supersonic jet expansion, flowing at 2000 sccm, that cools down a gas mixture to Titan-like temperature (150 K) and reduces the pressure to 22.5 Torr (3000 Pa). A plasma discharge generated in the stream of that expansion is then used as the energy source to induce the chemistry, resulting in the formation of more complex molecules and solid particles. The temperature remains low ($\sim$200 K) in the plasma \citep{2017Icar..289..214S}. The residence time of the gas in the plasma cavity (in yellow in Figure \ref{fig:facility}c), where chemistry occurs, is $\sim$3.5 $\mu$s. Solid particles (50--500 nm in size) produced in the plasma cavity are then carried past the electrodes in the gas expansion before being jet-deposited onto substrates placed $\sim$5 cm downstream of the electrodes. During deposition, the particles stack on top of each other and produce a deposit of hundreds of nm to a few $\mu$m in thickness. In the COSmIC setup, the reactant gases are accelerated to Mach number 8 \citep{2006FaDi..133..289B}, which could compact the grains. But the velocity of the reactant gases is not sufficient to break the grains, as demonstrated by a previous scanning electron microscope study of the deposited particles \citep{2017Icar..289..214S}.

To produce the four ARC tholin samples used in this study, four different gas mixtures were used: N$_2$:CH$_4$ (95:5), i.e., the same mixtures as for the UNI-UV and JHU-Plasma samples, but also N$_2$:CH$_4$:C$_2$H$_2$ (94.5:5:0.5), Ar:CH$_4$ (95:5), and Ar:CH$_4$:C$_2$H$_2$ (94.9:5:0.1). Ultra-high purity (99.9998\%) N$_2$, CH$_4$, and Ar cylinders, as well as pre-mixed N$_2$:C$_2$H$_2$ (99.99\%) and Ar:C$_2$H$_2$ (99.99\%) cylinders were used. Gases were flown through the PDN using three MFCs (Matheson MFC 1479A). The addition of a more complex precursor like acetylene (C$_2$H$_2$), into the nominal N$_2$:CH$_4$ gas mixture allowed investigating the effect of trace constituents on the material properties of tholins. The addition of C$_2$H$_2$ has been found to promote the formation of more complex molecules in the gas phase and larger particles in the solid phase \citep{2014Icar..243..325S,2017Icar..289..214S,2020ApJ...905...45S} in COSmIC. Using Ar-based gas mixtures allowed the examination of the impact of nitrogen on the material properties of the tholins. The sample deposition times varied from 5 hours (for mixtures with C$_2$H$_2$ that have a higher production rate) to 10--13 hours (for the mixtures without C$_2$H$_2$) as listed in Table \ref{table:tholin_samples}, in order to produce a thick enough layer of grains for subsequent contact angle measurements.

\subsection{Sample collection, transportation, and storage}
Prior to shipping, the UNI samples were collected and sealed with parafilm in a sample container in a dry N$_2$, oxygen-free glove box and then stored in a 4 $^\circ$C refrigerator; the JHU samples were collected and stored in a dry N$_2$, oxygen-free glove box ($<$0.1 ppm H$_2$O, $<$0.1 ppm O$_2$); and the ARC samples were collected in ambient air and then stored in a dark, continuously dry-air purged container with desiccant. All samples were sealed in small containers for transportation. After receiving the tholin samples, they were immediately unsealed and transferred to a vacuum desiccator that is continuously pumped down through a vacuum line. During this transfer, the samples were briefly exposed to air. However, because all seven tholin samples were stored in the vacuum desiccator for an extended period of time (days to weeks), it is considered that any water adsorbed on the surface would be desorbed from the surface prior to the contact angle measurements, as \citet{2020dis..xx} demonstrated that water desorbs from tholin surface under vacuum in around 30 minutes. By having the samples in a continuous vacuum prior to measurement, we are not only minimizing the air exposure and removing adsorbed water on the samples, but also reducing the differences between the surface properties due to their different collection and storage methods prior to shipping.
 
\subsection{Contact Angle Measurements}\label{subsec:sessile drop}
\begin{figure}[ht] 
    \centering
    \includegraphics[width=\textwidth]{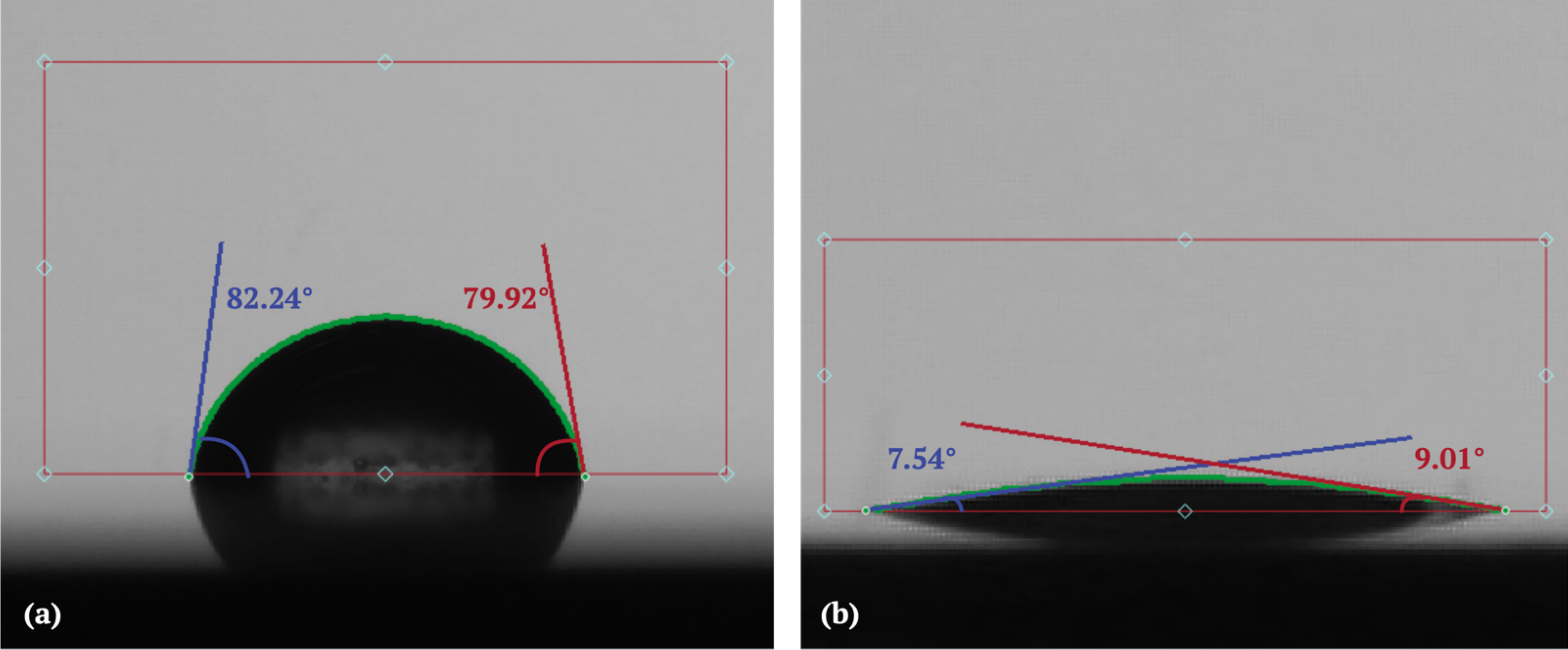}
    \caption{(a) Osilla's polynomial fit for acquiring the contact angle formed between water and the ARC-Plasma-Ar-1 tholin sample. The left and right contact angles are both above 10$^\circ$. Note that the average values of the left and right contact angles are used in the surface energy calculations. (b) Osilla's circle fit for the contact angle formed between water and the JHU-Plasma tholin sample with contact angle below 10$^\circ$.}
    \label{fig:fit}
\end{figure}

The first step in the contact angle measurement protocol was to place the vacuum desiccator inside a glove box (relative humidity RH$<1\%$) purged with 99.999$\%$ high-purity dry nitrogen. This allowed the samples to be extracted in a nitrogen atmosphere in order to avoid water adsorption, contamination, and surface aging in ambient air. All the contact angle measurements conducted on the samples were then performed inside the glove box. We first measured the surface roughness of the tholin samples with a Mitutoyo surface roughness tester to examine the smoothness of the tholin samples. We used a stylus with a small measuring force of 0.75 mN to prevent it from scratching the surface. All tholin samples and substrates were measured to be relatively smooth with root-mean-squared (RMS) roughness between 12--15 nm, measured with a cut-off length of 0.08 mm and a sample length of 0.4 mm.

The contact angles were then measured through the sessile drop method using two different test liquids: HPLC-grade water (Fisher Chemical\textsuperscript{TM}) and diiodomethane ($>$99$\%$, ACROS Organics\textsuperscript{TM}). The surface tensions of the two test liquids and their partitioning components are summarized in Table \ref{tab:testliquids}. The surface energy calculations are highly sensitive to the choice of the two test liquids, and the combination of a polar liquid and a non-polar liquid typically returns more reliable results \citep{2000JAPoSc...76...1831}. In particular, the water/diiodomethane pair tends to return the most accurate values as they have respectively the largest polar and non-polar components in all the test liquids that are commonly used \citep{2010wds3.conf...25H,2020ApJ...905...88Y}.

\begin{table}
\caption{Surface tensions and surface tension components (in mN$/$m) of the two test liquids used in this study at 20$^\circ$C, where $\gamma_{lv}^{tot}$ is the total surface tension of the test liquid, $\gamma^{d}_{lv}$ and $\gamma^{p}_{lv}$ are respectively the dispersive and polar components of the surface tension of the test liquids. The surface tension values are from \citet{2006VanOss}.}
 \begin{center}
 \begin{tabular}{c  c r c  r c c }
 \toprule
 Liquid & CAS number && Total surface tension && \multicolumn{2}{c}{Surface Tension Components} \\
\cmidrule{4-4} \cmidrule{6-7} 
& && $\gamma_{lv}^{tot}$ && $\gamma_{lv}^{d}$ & $\gamma_{lv}^{p}$\\
\midrule
Water &7732-18-5& & 72.8 && 21.8 & 51.0 \\
\hline
Diiodomethane & 75-11-6 & & 50.8 && 50.8 & 0\\
\bottomrule
 \end{tabular}
  \label{tab:testliquids}
  \end{center}
 \end{table}
 
Using a microsyringe, two to seven droplets of each test liquid were dispensed onto the tholin samples, forming sessile drops. The number of droplets we dispensed depends on the total available surface area of each sample. To avoid the flattening effect of gravity on the droplets, the volume of each droplet was controlled to be $<2\ \mu$L \citep{2005Langm...26..17090,2008JMChem...18..621}, forming sessile drops with diameters around 1--3 mm  on the samples. The sessile drops were recorded as 30--60 s videos, with a frame rate of 2--20 frames/s, using the Ossila goniometer software. Note that although each video was recorded for either 30 s or 60 s, the length of video containing contact angle data was typically shorter than the total length of video acquired. This is because the video recording needs to be started before dispensing the liquid droplet, and the operator often needs a few seconds to dispense the liquids from the syringe onto the tholin samples.

The static contact angles were measured by both the Ossila software of the goniometer and the contact angle plugin of the ImageJ software \citep{2012NMeth...9..671}. For the Ossila software, an adjustable box was used to enclose the region of interest. The base of the box was chosen to level the base of the droplet and an edge detection algorithm was used to find the edge of the droplet inside the box. A polynomial fit was used for contact angles that were above 10$^{\circ}$ as shown in Figure \ref{fig:fit}a, and a circle fit was used for contact angles that were below 10$^{\circ}$ as shown in Figure \ref{fig:fit}b. For the contact angle plugin of the ImageJ software, the baseline was defined via the selection of the left and right triphase points. The placement of three or more extra points along the droplet profile completed the definition of the droplet edge. The software fitted the profile to both a circle and an ellipse. Examples of the ImageJ fitting technique can be found in Figure 3 of \citet{2020ApJ...905...88Y}. 

We performed a comparison between the Osilla and ImageJ fits and confirmed that they returned similar contact angle values for a given droplet profile. For each tholin sample and for each test liquid, we compared around 10 random frames throughout a video that could be measured by both software packages. A total of about 100 individual frame comparison was performed, and the differences found between the two software packages were typically less than 5$^\circ$. Because of this relatively small difference, we use the two packages interchangeably.

\subsection{Surface Energy Derivation Methods}\label{subsec:SE}
The solid surface energies of the tholin samples ($\gamma_{sv}$, in mJ/m$^2$) can be obtained using the known surface tensions of the test liquids and the measured contact angles with the test liquids via the Young-Dupr\'e Equation \citep{1805RSPT...95...65,1869Dupre}: 
\begin{equation}\label{youngdupre}
W^{ad}_{sl}=\gamma^{tot}_{lv}(1+cos\theta),
\end{equation}
where $W^{ad}_{sl}$ is the work of adhesion between the test liquid and the solid or the free energy change to separate the liquid and solid phase per unit area, $\theta$ is the measured contact angle formed between the solid and the test liquid, and $\gamma^{tot}_{lv}$ is the total surface tension of the test liquid (in mN/m). The solid surface energy $\gamma_{sv}$ is included in the work of adhesion $W^{ad}_{sl}$.

The Young-Dupr\'e Equation enables the study of solid surface energy with the measurement of $\theta$ using a test liquid of known total surface tension $\gamma^{tot}_{lv}$. The surface tension component method \citep{1964I&EChem...56...40} can then be used to link the work of adhesion, $W^{ad}_{sl}$, to the solid surface energy, $\gamma_{sv}$. This theory assumes that the total surface free energy can be partitioned into different individual components representing contributions from different intermolecular forces, so W$^{ad}_{sl}$ can be expressed in terms of the surface free energy components.

In this study, we used the Owens-Wendt-Rabel-Kaelble (OWRK) method \citep{1969JAPSci...11..1741,1970JAdh...2...66,1971FL...77..997} which follows the Young-Dupr\'e theory to describe the relationship between $W^{ad}_{sl}$ and $\gamma^{tot}_{sv}$. This method, also known as the geometric mean method, assumes that the total surface energy ($\gamma^{tot}_{sv}$) or total surface tension ($\gamma^{tot}_{lv}$) can be partitioned into two parts: a dispersive component ($\gamma^d$) and a polar component ($\gamma^p$). Each part reflects independent intermolecular forces, with the dispersive component including the London dispersive forces between non-polar molecules, and the polar component including dipole-dipole and H-bonding interactions. The work of adhesion $W^{ad}_{sl}$ can be expressed as the sum of the geometric means of the surface free energy components:
\begin{equation}\label{owrk}
W^{ad}_{sl}=2(\sqrt{\gamma^{d}_{sv}\gamma^{d}_{lv}}+\sqrt{\gamma^{p}_{sv}\gamma^{p}_{lv}}),
\end{equation}
where $\gamma^{d}_{sv}$ and $\gamma^{p}_{sv}$ are the dispersive and polar components of the surface energy of the solid, whereas the $\gamma^{d}_{lv}$ and $\gamma^{p}_{lv}$ are the dispersive and polar components of the surface tension of the liquid.

Combining Equations \ref{youngdupre} and \ref{owrk}, we have:
\begin{equation}\label{owrk2}
 \gamma^{tot}_{lv}(1+cos\theta)=2(\sqrt{\gamma^{d}_{sv}\gamma^{d}_{lv}}+\sqrt{\gamma^{p}_{sv}\gamma^{p}_{lv}}).
\end{equation}
 Due to the partitioning of $\gamma^{tot}_{sv}$ into $\gamma^{d}_{sv}$ and $\gamma^{p}_{sv}$ in Equation \ref{owrk2}, two unknowns are present. Therefore, contact angles measured between two known liquids and the solid are needed to form two sets of Equation \ref{owrk2} in order to solve for the total surface energy of the solid ($\gamma^{tot}_{sv}$) and its partitioning components ($\gamma^{d}_{sv}$ and $\gamma^{p}_{sv}$). The analytical solution of the OWRK method and the 1$\sigma$ standard deviations for $\gamma^{tot}_{sv}$, $\gamma^{d}_{sv}$, and $\gamma^{p}_{sv}$, computed through the propagation of uncertainty, can be found in the Appendix Section.

\section{Results and Discussion}

\subsection{Contact Angles Between Tholins and Test Liquids}

\begin{figure}[ht] 
    \centering
    \includegraphics[width=\textwidth]{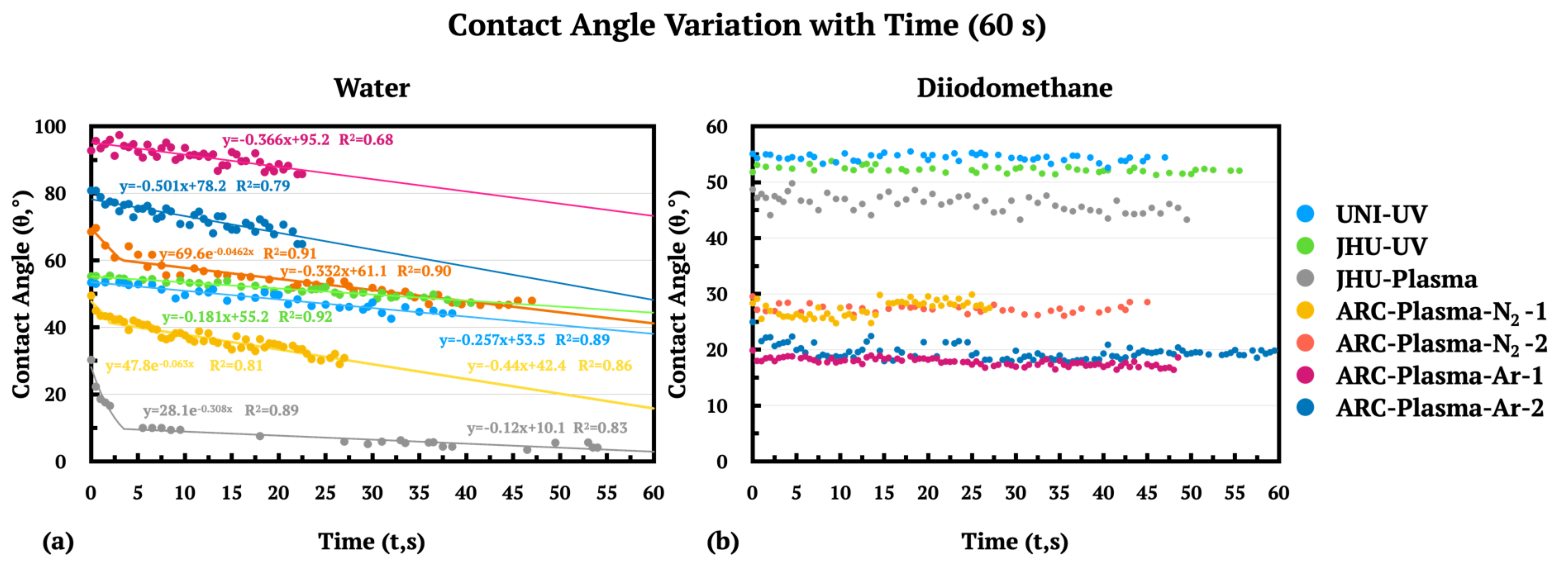}
    \caption{Contact angle variation with time for all tholin samples after the immediate formation of a sessile drop on the surfaces of the tholin samples for one selected water droplet (left) and one selected diiodomethane droplet (right) per sample. We fit the contact angle variation with time data with a combination of exponential and linear fits for water. The fitting functions along with their corresponding R$^{2}$ values are shown for each tholin sample. Exponential fittings are only performed for samples with a dissolution driven decay (JHU-Plasma, ARC-Plasma-N$_2$-1, and ARC-Plasma-N$_2$-2). ARC-Plasma-N$_2$-1, -2 stand for ARC-Plasma-N$_2$:CH$_4$ (95:5) and ARC-Plasma-N$_2$:CH$_4$:C$_2$H$_2$ (94.5:5:0.5), respectively; ARC-Plasma-Ar-1, -2 stand for ARC-Plasma-Ar:CH$_4$ (95:5) and ARC-Plasma-Ar:CH$_4$:C$_2$H$_2$ (94.9:5:0.1), respectively.}
    \label{fig:60s}
\end{figure}

Figures \ref{fig:60s}a and \ref{fig:60s}b show one set of measured contact angles formed between each of the seven tholin samples and single droplets of each of the two test liquids,  water and diiodomethane, as well as their variation with time. The whole contact angle dataset for all droplets can be found in a supplementary data file in a Dryad repository available at \dataset[10.7291/D13T16]{https://doi.org/10.7291/D13T16}. The contact angles were measured by either the Ossila software or the ImageJ contact angle plugin as described above. Most contact angles were measured by the Ossila software, however, because the automatic edge detection algorithm is extremely sensitive to the defects of the background enclosed in the area of interest (see Figure \ref{fig:fit}): if a defect in the background (such as another droplet) is detected, no result is returned. For those cases with defects, the contact angles were instead measured using the ImageJ contact angle plugin.

With water as the test liquid, all tholin samples have decreasing contact angles over time, as shown in Figure \ref{fig:60s}a. The decrease in contact angles for diiodomethane is more subtle during the $\lesssim$ 60 s video recording period and is only visible when the contact angle variation is recorded over an extended amount of time, an example is shown in Figure \ref{fig:extended} in the Appendix B section. The decay of contact angle with time could be caused by the spreading of the liquid droplet, penetration of the test liquid into pore spaces, dissolution of the solid material into the test liquid, and evaporation of the test liquid to the measuring atmosphere. \citet{2020CI...4...48} classified the process of contact angle decay into three stages. The spreading kinetics dominates the first stage, which lasts for only tens of ms and is usually hard to be experimentally measured. This stage is dependent on the porosity of the material \citep{2002langmuir..xxx} and independent of the solubility of the material. At the second stage, spreading starts to depend on the solubility of the solid material in the test liquid. This stage typically lasts for a few seconds, and higher solubility would lead to shorter decay timescales. The contact angle decays exponentially in this stage and higher solubility leads to more rapid decay. The third stage starts after the contact lines are pinned. At the third stage, evaporation dominates the mass loss of the droplet and the contact angle decays linearly with time until the droplet is completely evaporated.

Our experimental data (as shown in Figure \ref{fig:60s}) do not have the resolution (0.05--0.5 s per data point) to reflect the first stage of spreading. However, we were able to fit the data with a combination of exponential and linear fits to examine the effect of dissolution and evaporation. Consistent with the theory described by \citet{2020CI...4...48}, we found that dissolution typically occurred in the first few seconds after the sessile drop was placed onto the tholin sample, while evaporation happened later on. Dissolution also led to sharper decreases of contact angles while the decrease due to evaporation was more gradual. Among all the tholin samples, only three samples are strongly affected by dissolution, the JHU-plasma sample, the ARC-Plasma-N$_2$-1 sample, and the ARC-Plasma-N$_2$-2 sample. We are able to fit an exponential decay function to the first few seconds of the data and a linear fit for the remaining data in Figure \ref{fig:60s}a for these three samples. The JHU sample is known to be partially soluble in water \citep{2020ApJ...905...88Y}. The ARC-Plasma-N$_2$ samples do not have measured solubility, but their contact angle decay data suggests that the ARC tholin samples produced with cold plasma discharge in N$_2$-based gas mixtures are also soluble in water. Among the three samples that show signs of dissolution with water, the JHU-plasma sample has the fastest dissolution-driven decay (highest slope in the exponential fit) and thus has the highest solubility. For the rest of the tholin samples, a linear fit was sufficient for capturing the contact angle decay data with water (Figure \ref{fig:60s}a). Unlike their N$_2$-based counterparts, the tholin samples produced in Ar-based mixtures, ARC-Plasma-Ar-1 and ARC-Plasma-Ar-2, had much slower decline rates in their contact angles with water in the first few seconds, indicating that they are much less soluble or even insoluble in water. This insolubility is likely attributed to the polarity of these samples. Water is more likely to dissolve polar materials, such that samples containing nitrogen more readily dissolve in water (see more detailed discussion in the following section). The gas mixtures used to produce the ARC-Plasma-Ar samples only contained non-polar elements, so it is expected that the resulting solid samples are largely non-polar as well. Therefore, it is reasonable to conclude that the ARC samples produced in Ar-based gas mixtures are insoluble in water because they are non-polar. Also, both the JHU-UV and UNI-UV tholin samples, which were produced from the same initial gas mixture as the JHU-Plasma and ARC-Plasma-N$_2$-1 tholin samples, have little solubility or are insoluble in water. This is likely due to the fact that they were produced using UV lamps as the energy source to induce the chemistry and that UV lamps cannot directly dissociate nitrogen. Indeed, the UV lamps used in the PAC and PHAZER experimental setups emit photons from 115 to 400 nm, while the wavelength required for direct photolysis of nitrogen has to be shorter than 80 nm \citep{1981JGR....86.1495R}. It is therefore expected that less nitrogen is incorporated in these UV-generated tholin samples, resulting in compounds that have minimal solubility in water.

Tholins are typically found to be much less soluble in non-polar solvents than polar solvents \citep{1996P&SS...44..741M,2009JPCA..11311195C,2013IJAsB..12..282K,2014Icar..238...86H,2020PSJ.....1...17M}. As shown in Figure \ref{fig:60s}b, clearly, none of the samples dissolve in diiodomethane as there is no sharp decay of contact angles in the first few seconds of the data. We attempted to fit the 60 s contact angle decay data with linear fits but were unable to find good correlations (coefficient of determination R$^2<0.6$) due to the scattering of the data. We recorded an extended contact angle decay for the JHU-plasma sample and we are able to fit the data with a linear fit with a more significant R$^2=0.83$. The slope of the linear decay for diiodomethane ($\sim$0.02 $^\circ/s$) is much smaller than that of water ($\sim$0.1--0.5 $^\circ/s$). This could be due to diiodomethane's higher boiling point ($\sim$180$^\circ$C) than water ($\sim$100$^\circ$C) and thus lower vapor pressure at room temperature, so water more readily evaporates in a dry nitrogen environment than diiodomethane, leading to a more drastic contact angle decrease.

To accurately capture the surface properties of the tholin samples, we wanted to record the contact angles shortly after the first stage of spreading and minimize the effects of dissolution and evaporation. For diiodomethane, we averaged the contact angles measured for the full 60 s period (Table \ref{table:contact_angles}), as the sessile drop was relatively stable with little change of the contact angles due to evaporation and/or dissolution, as shown in Figure \ref{fig:60s}b. For the water droplet measurements, only the contact angle values from the first 5 s after the water droplet was dispensed on the surfaces of tholins were taken into consideration for all samples, to minimize the effect of water evaporation. In the case of the three plasma tholin samples (JHU-Plasma, ARC-Plasma-N$_2$-1, and ARC-Plasma-N$_2$-2) that partially dissolve in water, this time was even shorter to reduce the effect of dissolution in water, and only the contact angle values from the first 2.5 s after the dispensary of droplets were taken into account. For each droplet, we averaged the contact angles measured from all selected frames and calculated the 1$\sigma$ contact angle standard deviation due to frame-to-frame variations. We then averaged the contact angle values for all the droplets to obtain the final contact angle value for each sample with each test liquid and propagate contact angle error using the average and 1$\sigma$ contact angle standard deviation values of each droplet. The final averaged contact angles and their 1$\sigma$ standard deviations between each tholin sample and the two test liquids are summarized in Table \ref{table:contact_angles}.

Ideally, the contact angle measurement would most accurately probe for the surface property of the tholin samples when they have perfectly smooth surfaces with no voids (i.e., by measuring the ``intrinsic" contact angles). The existence of surface roughness and porosity could both lead the measured contact angles (or the ``apparent" contact angle) to deviate from the intrinsic contact angles. All seven tholin samples used in this study were measured to be very smooth, with RMS roughness of 12--15 nm on a scale of 0.08 mm. Such small surface roughness indicates low porosity (less than a few percent) as well, as these two properties are linearly correlated \citep{2015AcMat..96...72Q}. Previous microscopic images of the JHU-plasma tholin sample used here \citep{2017JGRE..122.2610Y} also show that its surface is densely packed with few voids. Given that our contact angle measurements already have standard deviations around $\sim$5$^\circ$ for each liquid on each sample due to frame-to-frame variations and variations between droplets at different locations (see Table \ref{table:contact_angles}), the effect of the porosity and surface roughness would be minimal compared to the measurement error. Thus, the apparent contact angles measured for the tholin samples here should be close to their intrinsic contact angles and the subsequent surface energy calculations likely capture the intrinsic surface energy of the tholin samples.

\begin{table}
\caption{Measured contact angles between the tholin samples and the two test liquids, and their respective 1$\sigma$ standard deviations.}
\label{table:contact_angles}
\centering
\begin{tabular}{l cc r cc}\toprule
 \multirow{3}{*}{Sample} &  \multicolumn{5}{c}{Test Liquids}\\
 \cmidrule{2-6}
 &  \multicolumn{2}{c}{Water} && \multicolumn{2}{c}{Diiodomethane} \\
 \cmidrule{2-3}  \cmidrule{5-6}
 & Contact Angle ($^\circ$) & StDev ($^\circ$)  &&  Contact Angle ($^\circ$) & StDev ($^\circ$) \\
\midrule
UNI-UV& 58.2 & 9.3&& 50.7 &5.2\\

JHU-UV & 60.8 & 4.6&& 54.4 & 3.1  \\
JHU-Plasma & 16.2 & 4.0&& 44.4 & 2.9\\

ARC-Plasma-N$_2$-1 & 41.7& 2.9 && 26.3 & 5.6\\
ARC-Plasma-N$_2$-2 & 63.6& 10.7 && 29.6& 2.4 \\
ARC-Plasma-Ar-1 & 89.4&1.9 && 18.0&0.6\\
ARC-Plasma-Ar-2 & 81.9&4.4 && 21.0&2.1\\

\bottomrule
\end{tabular}
\end{table}

\subsection{Surface Energy of Tholins}

The surface energies of the tholin samples were computed using the OWRK method described in Section \ref{subsec:SE} with the contact angle values from Table \ref{table:contact_angles} and the known surface tensions of the test liquids in Table \ref{tab:testliquids}. We also calculated the surface energies of the samples using the Wu method described in \citet{2020ApJ...905...88Y} as a way to validate our results. The Wu method returned slightly higher total surface energy values than the OWRK method, but the deviation in the values obtained by the two methods was only a few mJ/m$^2$. The total surface energies, $\gamma^{tot}_{sv}$, and their partitioning components calculated with the OWRK method are listed in Table \ref{table:SE}. The resulting total surface energies, spanning from 47.0--73.3 mJ/m$^2$, are similar or higher than the high end of surface energies obtained for common polymers made of C, N, and H (20--50 mJ/m$^2$; \citealt{1969JAPSci...11..1741,1971JPoSc...34...19,1982Wu}). Thus, the tholins studied here are stickier than common polymers and other materials that have been used in past studies to simulate Titan sediments \citep{2015Natur.517...60B, 2017NatGe..10..260M}. If the surface sediments on Titan resemble the tholins studied here, then they are likely relatively sticky, which would have an impact on their transport on the surface of Titan \citep{2021xxxcomola}.

Out of all the tholin samples, the JHU-Plasma sample has the highest overall surface energy. Even though the two JHU samples (JHU-Plasma and JHU-UV) were produced under very similar experimental conditions (same pressure, temperature, gas mixture) other than their energy sources, the differences in their surface energies are significant. Both the JHU-plasma and JHU-UV samples have a relatively large dispersive component (over 30 mJ/m$^2$), but the JHU-plasma sample has a much higher polar component compared to the JHU-UV sample. The smaller polar component of the JHU-UV sample can be attributed to the hydrogen UV lamp's inability to directly dissociate the triple-bonded nitrogen in the gas mixture. Because nitrogen-containing compounds are the main contributors to the polar components of the tholin samples, the lower polar component observed with the JHU-UV sample can therefore be explained by the lack of directly dissociated molecular nitrogen during the JHU-UV sample production with the UV lamp. The non-zero polar component of the JHU-UV sample however is consistent with the fact that nitrogen can still be indirectly dissociated through various secondary processes \citep{2011Icar..214..748H, 2012AsBio..12..315T, 2018Icar..301..136H} and incorporated in the tholin sample. 

\begin{table}
\caption{Derived surface energies, $\gamma_{tot}^{sv}$, and their two partitioning components, $\gamma_{sv}^{d}$ and $\gamma_{sv}^{p}$, with their respective 1$\sigma$ standard deviation in mJ/m$^2$ for all seven tholin samples studied as well as the ``average" and ``end-member" surface energies of tholins.}
\label{table:SE}
\centering
\begin{tabular}{c  crcrc}\toprule
 Sample &  \multicolumn{5}{c}{OWRK Method} \\
 \cmidrule{2-6}
 &  $\gamma_{sv}^{d}$ &&  $\gamma_{sv}^{p}$ && $\gamma_{tot}^{sv}$ \\
\midrule
UNI-UV & 33.9 $\pm$ 2.9 && 15.8 $\pm$ 5.7 && 49.7 $\pm$ 5.8 \\
JHU-UV & 31.8 $\pm$ 1.8 && 15.2 $\pm$ 2.9 && 47.0 $\pm$ 2.9\\
JHU-Plasma & 37.3 $\pm$ 1.5 && 36.0 $\pm$ 1.5 && 73.3 $\pm$ 1.3\\
ARC-Plasma-N$_2$-1 & 45.7 $\pm$ 2.1 && 20.1 $\pm$ 1.8 && 65.8 $\pm$ 1.9 \\
ARC-Plasma-N$_2$-2 & 44.4 $\pm$ 1.0 && 9.0 $\pm$ 5.1 && 53.4 $\pm$ 5.2 \\
ARC-Plasma-Ar-1 & 48.3 $\pm$ 0.2 && 0.4 $\pm$ 0.2 && 48.7 $\pm$ 0.3\\
ARC-Plasma-Ar-2 & 47.5 $\pm$ 0.6 && 1.7 $\pm$ 1.0 && 49.2 $\pm$ 1.2 \\
\midrule
Average surface energy & 37.2 && 21.8 && 59.0\\
End-member surface energy--low $\gamma_{sv}$& 31.8 && 15.2 && 47.0 \\
End-member surface energy--high $\gamma_{sv}$& 45.7 && 36.0 && 81.7\\
\bottomrule
\end{tabular}
\end{table}

In a previous study of the same JHU tholin samples by \citet{2020ApJ...905...88Y}, closer surface energy values between the JHU-plasma and JHU-UV samples were reported. The values, obtained using the two-liquid method, ranged from 60--70 mJ/m$^2$ for both samples. The surface energy value reported for the JHU-Plasma sample was close to the one obtained in this study. However, the surface energy value reported for the JHU-UV sample in Yu et al. (2020, $\gamma_{sv}^{tot} = 66.0\pm4.7\ mJ/m^2$, $\gamma_{sv}^d = 41.1\pm0.9\ mJ/m^2$, $\gamma_{sv}^p = 24.9\pm4.7\ mJ/m^2$) was much higher than the value obtained in this study ($\gamma_{sv}^{tot} = 47.0\pm2.9\ mJ/m^2$, $\gamma_{sv}^d = 31.8\pm1.8\ mJ/m^2$, $\gamma_{sv}^p = 15.2\pm2.9\ mJ/m^2$). The JHU-UV sample used here and the one used in \citet{2020ApJ...905...88Y} are the same batch of samples that have been stored in dry nitrogen, oxygen-free glove box to prevent sample aging. Thus, the most likely cause of the discrepancy comes from the different measurement environments: ambient air \citep{2020ApJ...905...88Y} versus dry nitrogen in this study. As discussed in \citet{2020ApJ...905...88Y}, the surface properties of the tholin samples could be affected in several different ways when the contact angle measurements were performed in ambient air: 1) the tholin samples could be chemically altered upon exposure to oxygen and water; 2) water moisture from the atmosphere may adsorb on the surfaces and steer the surface energy of the samples towards water (72.8 mN/m) ; 3) airborne organic contaminants like hydrocarbons/oils could also adsorb on the surfaces and steer the surface energy of the samples towards hydrocarbons/oils ($\sim$20 mN/m). For the JHU-UV sample, the water adsorption seems to be more significant than hydrocarbon adsorption in \citet{2020ApJ...905...88Y}, since the JHU-UV sample has lower surface energies measured here in an inert environment compared to ambient air. It is reasonable to expect the difference between the measurements in ambient air versus dry nitrogen to be smaller for the JHU-plasma sample than the JHU-UV sample, because the JHU-plasma sample (in dry nitrogen) has a surface energy ($\sim$70 mJ/m$^2$) that is closer to water (72.8 mN/m) than the JHU-UV sample ($\sim$50 mJ/m2), so the adsorbed water likely has less effect on the JHU-plasma sample. Thus, the discrepancy between this study and \citet{2020ApJ...905...88Y} demonstrates the need to perform the contact angle measurements in inert atmospheres to accurately capture the surface properties of tholin samples to avoid adsorption and contamination from ambient air.

If we now compare the UNI-UV and JHU-UV samples, which were made with two completely different experimental setups under different pressure and temperature experimental conditions but similar energy sources and the same gas mixture, we can see that they have similar overall surface energies and surface energy partitioning patterns. This resemblance between the two samples could be attributed to the common experimental parameters used during their production: the energy source, a UV lamp, and the initial gas mixture, N$_2$:CH$_4$ (95:5). The UV lamps used for the production of the UNI-UV and JHU-UV samples emit photons at similar wavelength ranges (115--400 nm), which means neither emits the range of photons that directly dissociates nitrogen in the gas mixture, as discussed above. The amount of polar molecules formed during the production of both these two samples was therefore lower compared to tholins formed with plasma discharges, which can dissociate nitrogen. The similarity of total surface energies of both UV samples is thus likely attributable to the energy source used to produce them.

The ARC samples all have very high dispersive components ($>$ 44 mJ/m$^2$) that are higher than the JHU and UNI samples. The surface energies of the ARC samples made with different gas mixtures differ mainly in their polar components. The ARC-Plasma-N$_2$-1 sample, made with the same gas mixture as the UNI and JHU samples, has a surface energy polar component that is higher than that of the two UV samples but lower than that of the JHU-Plasma sample. For both the ARC-Plasma-N$_2$-1 and JHU-Plasma samples, the use of a plasma discharge as the energy source allowed for the dissociation of molecular nitrogen during production of tholins. A possible explanation for the higher polar component of the JHU-Plasma sample's surface energy compared to that of the ARC-Plasma-N$_2$-1 sample is the exposure time of the gas mixture to the energy source: $\sim$3.5 $\mu$s for the ARC-Plasma-N$_2$-1 sample versus 3 min for the JHU-Plasma sample, i.e. $\sim$50,000 times shorter for the ARC sample. Note that the ARC samples are also collected and stored differently compared to the JHU-plasma sample (see Section 2.2), which could cause different degrees of sample degradation. Some of the differences in collection and storage could be reduced by the prolonged storage of samples under a continuous vacuum prior to measurements, such as water and hydrocarbon adsorption. However, chemical alteration in air is irreversible and could also contribute to the differences between the ARC and JHU samples. On the other hand, the higher polar component of the surface energy observed for the plasma-generated ARC-Plasma-N$_2$-1 tholin sample compared to the UV-generated PAC and JHU tholin samples is in agreement with our previous assessment that direct dissociation of nitrogen must play an important role in forming molecules with high polarity indices in tholins. Given that the two UV samples (JHU-UV and UNI-UV) have similar surface energy values and partitioning values, and both the plasma samples (JHU-Plasma and ARC-Plasma-N$_2$-1) have higher surface energies than the UV samples, it seems like the energy source has the largest impact on the surface energies or the bulk chemical compositions of the tholin samples, which is in agreement with a previous study by \citet{2018Icar..301..136H}.

A previous mass spectrometry study of the plasma-generated gaseous chemical products formed in COSmIC, where the ARC tholin samples were produced, has shown that mixtures containing heavier precursors, like N$_2$:CH$_4$:C$_2$H$_2$, produce more complex molecular compounds than simpler gas mixtures like N$_2$:CH$_4$ \citep{2014Icar..243..325S}. It is interesting to observe however that the surface energies of the tholins formed in N$_2$:CH$_4$:C$_2$H$_2$ are not necessarily higher than those of tholins formed in simpler N$_2$-CH$_4$ gas mixtures. The ARC-Plasma-N$_2$-1 tholin sample produced without C$_2$H$_2$ has a much higher surface energy polar component than the ARC-Plasma-N$_2$-2 tholin sample produced with C$_2$H$_2$ (20 mJ/m$^2$ versus 9 mJ/m$^2$). This result suggests that the addition of a more complex non-polar hydrocarbon precursor in the initial gas mixture leads to the formation of a tholin sample that contains more complex non-polar products in COSmIC. This is consistent with a recent X-ray Absorption Near Edge Structure (XANES) spectroscopy study of COSmIC tholins that allowed the measurement and comparison of the C/N ratios of the tholin samples, and demonstrated that COSmIC tholin samples produced in N$_2$:CH$_4$ (95:5) had a higher nitrogen content than the ones produced in N$_2$:CH$_4$:C$_2$H$_2$ (94.5:5:0.5) (1.3 versus 2.4, \citealt{2021LPI....52.2160N}). 

\begin{figure}[ht] 
    \centering
    \includegraphics[width=\textwidth]{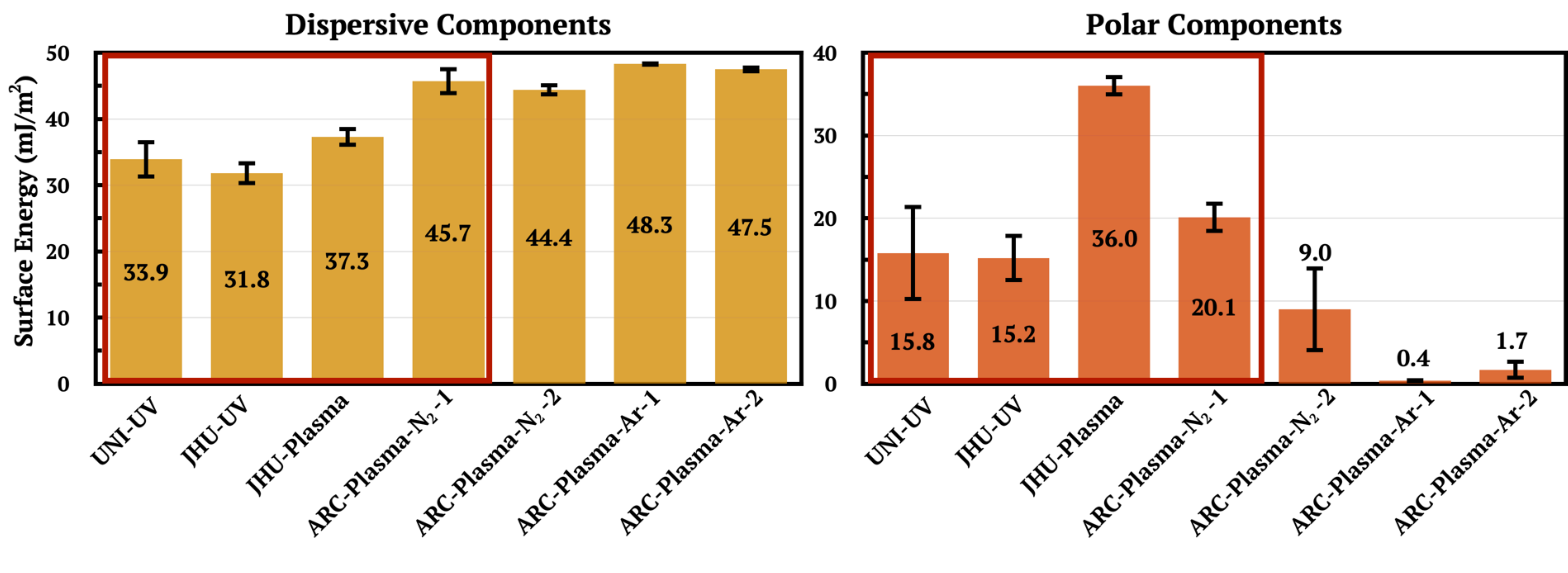}
    \caption{Surface energy components of the seven tholin samples along with their associated 1$\sigma$ standard deviations, calculated using the OWRK method. Left: calculated dispersive components ($\gamma_{sv}^{d}$) (in yellow). Right: calculated polar components ($\gamma_{sv}^{p}$) (in orange). The four samples produced in the N$_2$:CH$_4$ (95:5) mixture more representative of Titan's atmospheric composition are highlighted in the red boxes.}
    \label{fig:diffcomp}
\end{figure}

The ARC-Plasma-Ar-1 and ARC-Plasma-Ar-2 samples have very similar total surface energies and surface energy partitioning components, as seen in Figure \ref{fig:diffcomp}. The polar components of the surface energy for both the ARC-Plasma-Ar-1 and ARC-Plasma-Ar-2 samples are almost 0 mJ/m$^2$. Because the initial gas mixtures for these two samples contained only non-polar constituents, these two samples are likely composed of only non-polar compounds. This result confirms that nitrogen indeed plays a prominent role in the formation of polar compounds for the tholins studied here.

\subsection{Commonality and End Member Surface Energies of Tholins}
\label{common}
From the measured surface energies of the different tholin samples, we can extract their commonalities and end member properties. These can then be used to infer what properties could be expected to be present in Titan's aerosols, and to estimate the processes that are likely happening on Titan. This end member selection method was previously used to determine the range of possible optical properties of tholins to compare DISR and VIMS observations \citep{2006Icar..185..301B}. Here we considered only the four tholins produced in the same Titan-representative gas mixture that is commonly used in the literature, N$_2$:CH$_4$ (95:5), i.e. UNI-UV, JHU-UV, JHU-Plasma, and ARC-Plasma-N$_2$-1. These four samples are selected to balance the effect of each apparatus and each energy source, such that two plasma tholin samples and two UV samples from two different laboratories go into the decision of the ``average" and the ``end-member" properties.

As shown in Figure \ref{fig:diffcomp}, the four tholin samples produced in N$_2$:CH$_4$ (95:5) gas mixtures all have relatively high dispersive components above 30 mJ/m$^2$. This suggests that a high dispersive component ($\gamma^d_{sv}$) may be a common trait among all tholin samples. Thus, the actual Titan haze particles could likely also share this trait and have a high dispersive component ($\gamma^d_{sv}$). The differences between the tholin samples are mainly their polar components ($\gamma^p_{sv}$), which range from 15 to 36 mJ/m$^2$. Overall, cold plasma energy sources seem to produce tholin samples with higher polar components than UV energy sources. However, differences in other experimental conditions could also lead to the observed differences in the polar components of the tholin samples.

We have defined an ``average" and two ``end-member" surface energies, as shown in Table \ref{table:SE}. The dispersive and polar components of the average surface energy were obtained by averaging the dispersive and polar components, respectively, of the four Titan tholins produced in N$_2$:CH$_4$ (95:5, UNI-UV, JHU-UV, JHU-Plasma, and ARC-Plasma-N$_2$-1). Note that the average dispersive surface energy component also represents the common trait of all the tholin samples. The surface energy partitioning components of the end-member surface energies correspond to the end member dispersive and polar components, i.e., the highest and lowest possible values, of the four Titan tholin samples.

\subsection{Application to processes on Titan}

\begin{table} 
\caption{Surface tensions of Titan's lake liquids (methane, ethane, and nitrogen) in mN/m, and surface energies of Titan's methane and ethane ice cloud condensates, adopted from \citet{2020ApJ...905...88Y}.}
 \centering
 \begin{tabular}{c  c  c c r crcc}\toprule
 Species & \multicolumn{3}{c}{Liquid at 94 K} && \multicolumn{4}{c}{Solid $<$ 90 K}\\
 \cmidrule{2-4}  \cmidrule{6-9}
 & Total surface tension& \multicolumn{2}{c}{ Surface tension components}  && Total surface energy && \multicolumn{2}{c}{Surface energy components}\\
 & $\gamma_{lv}^{tot}$ & $\gamma_{lv}^d$ & $\gamma_{lv}^p$  && $\gamma_{sv}^{tot}$ && $\gamma_{sv}^d$ & $\gamma_{sv}^p$\\
\midrule
Methane & 17.8 & 17.8 & 0 && 23.5 && 23.5 & 0\\
Ethane & 31.8 & 31.8 & 0 && 43.4 && 43.4 & 0 \\
Nitrogen & 5.3 & 5.3 & 0 && N/A && N/A & N/A \\
\bottomrule
\end{tabular}
\label{table:hydrocarbons}
\end{table}
 
If the surface energy of Titan haze particles were known, we could determine two processes happening on Titan: cloud formation through haze-condensate interactions and haze-lake interactions \citep{2020ApJ...905...88Y}. In Section \ref{common}, we determined the ``average" and ``end member" surface energies of the tholin samples studied here. We used these surface energies as inputs in theoretical calculations to estimate a range of cloud formation and haze-lake interaction scenarios on Titan.

In Titan's atmosphere, simple organics such as methane and ethane can condense and form clouds (e.g., \citealt{2017P&SS..137...20B,2018SSRv..214..125A}). Clouds can form through both heterogeneous and homogeneous nucleation with the condensation of the organic species. However, heterogeneous nucleation is typically more favorable as it usually requires a lower degree of supersaturation \citep{1978Pruppacher}. Because of the prevalence of haze particles in Titan's atmosphere, they are natural potential CCN for heterogeneous nucleation. Thus, determining haze-condensate interactions on Titan can tell us the cloud forming efficiency for certain condensates and the likelihood of observing these clouds in Titan's atmosphere.

Using the surface energy of the Titan tholins as well as the average and end-member surface energies, along with the surface free energies for methane and ethane condensates, the theoretical contact angles with these hydrocarbon condensates can be calculated using the following equation \citep{2020ApJ...905...88Y}:
\begin{equation}
\begin{aligned}\label{eq:thetacalc_0}
\textrm{when} \ \sqrt{\gamma^d_{sv}\gamma^d_{lv}}+\sqrt{\gamma^p_{sv}\gamma^p_{lv}}\leq\gamma_{lv}^{tot},
\theta=arccos(\frac{2(\sqrt{\gamma^d_{sv}\gamma^d_{lv}}+\sqrt{\gamma^p_{sv}\gamma^p_{lv}})}{\gamma_{lv}^{tot}}-1),\\
\textrm{when} \ \sqrt{\gamma^d_{sv}\gamma^d_{lv}}+\sqrt{\gamma^p_{sv}\gamma^p_{lv}}\geq\gamma_{lv}^{tot}, \theta=0,
\end{aligned}
\end{equation}
where $\gamma_{lv}^{tot}$, $\gamma_{lv}^d$, and $\gamma_{lv}^p$ represent respectively the total surface free energy and the dispersive and polar components of the surface free energy (surface tension/surface energy) of cloud condensates, $\gamma_{sv}^{tot}$, $\gamma_{sv}^d$, and $\gamma_{sv}^p$ represent the total surface energy and the dispersive and polar components of tholins. Since the cloud condensates studied here (methane, ethane) and the lake liquid components (nitrogen, methane, ethane) are all non-polar in nature, their polar parts of the surface free energy are zero ($\gamma_{lv}^p=0$), and their dispersive and total surface free energies are equal to each other ($\gamma_{lv}^d=\gamma_{lv}^{tot}$). Thus Equation \ref{eq:thetacalc_0} can be simplified as:
\begin{equation}
\begin{aligned}\label{eq:thetacalc}
\textrm{when} \ \gamma^d_{sv}\leq\gamma^d_{lv},
\theta=arccos(2\sqrt{\frac{\gamma^d_{sv}}{\gamma^d_{lv}}}-1),\\
\textrm{when} \ \gamma^d_{sv}\leq\gamma^d_{lv}, \theta=0.
\end{aligned}
\end{equation}
Equation \ref{eq:thetacalc} has two parts. When $\gamma^d_{sv}\leq\gamma^d_{lv}$, the condensate will form a finite contact angle on the surface of the tholin sample. When $\gamma^d_{sv}\geq\gamma^d_{lv}$, the first part of Equation \ref{eq:thetacalc} does not apply anymore, as once the surface energy of the tholin sample is larger than the condensate, the condensate will completely wet (or spread on) the solid surface \citep{1950xxx..12..315C,1952xxx..12..315C}, forming a contact angle $\theta$ of zero. \citet{2020ApJ...905...88Y} has also validated the second part of this theory on tholins, by dispensing droplets of n-hexane ($\gamma^d_{lv}=18.4$ mN/m) on the JHU tholin samples (both have higher $\gamma^d_{sv}$), they form contact angles of zero on both samples. Based on Equation \ref{eq:thetacalc}, the dispersive component of the haze particles solely determines the interaction between the haze and the cloud condensates/lake liquids. Thus, using the ``average" surface energy determined in Section \ref{common} for the following calculations essentially corresponds to using the commonality in surface energy of the tholin samples studied here, which would likely be representative of the surface energy of Titan's aerosols. On the other hand, the modeling results obtained when considering the end member surface energies provide the upper and lower limits of a range of scenarios that could happen on Titan.

Considering the surface energy values for the main clouds, methane and ethane from Table \ref{table:hydrocarbons}, and the surface energy values provided in Table \ref{table:SE}, we have calculated theoretical contact angles for all four Titan tholin samples produced in N$_2$:CH$_4$ (95:5) gas mixtures as well as the ``average" and ``end member" surface energies defined above. The results are shown in Table \ref{table:predicted angles}. For the tholin samples, the standard deviations of the calculated contact angles are determined through error propagation using the 1$\sigma$ standard deviations of the surface energies of the tholin samples in Table \ref{table:SE}. When the calculations need to use both parts of Equation \ref{eq:thetacalc}, we calculate a range of $\theta$ values using upper and lower limit values of tholins' surface energies (surface energy plus/minus 1$\sigma$ standard deviation).

Calculations with all four Titan tholin samples result in 0$^\circ$ contact angle with liquid methane and methane ice. For liquid ethane, the JHU-UV and UNI-UV samples have contact angles ranging from 0--20$^\circ$ considering their measurement standard deviations. While for ethane ice, the contact angles range from 0$^\circ$ to 45$^\circ$. For the ``average" surface energy, a contact angle of 0$^\circ$ is obtained with liquid methane/ethane and methane ice, and a contact angle of 31.7$^\circ$ is obtained with ethane ice. For the ``end member" surface energy with the highest total surface energy $\gamma_{sv}$, 0$^\circ$ contact angles are obtained with methane and ethane in both liquid and solid phases, while for the ``end member" surface energy with the lowest surface energy $\gamma_{sv}$, a 0$^\circ$ contact angle is obtained with liquid methane/ethane and methane ice, but a contact angle of 44.6$^\circ$ is obtained with ethane ice. For a material to be an ideal CCN for efficient heterogeneous nucleation to form a cloud, the condensate has to have low contact angle with this material. Previous studies have demonstrated that the upper limit of contact angle values for good CCN ranges from 12$^\circ$ to 45$^\circ$ \citep{1964JAtS...21..109M,1975JAtS...32..116M,2021natastroxxx}. Thus, all tholins studied here should be good CCN for methane and ethane clouds. Assuming the haze particles are similar to the tholins studied here, the haze particles will likely be good CCN for methane and ethane clouds on Titan. There are other species that can condense in Titan's atmosphere, such as HCN, C$_6$H$_6$, and HC$_3$N \citep{2018SSRv..214..125A}, and the study of additional possible condensates is ongoing.

As the haze particles on Titan grow bigger and fall towards Titan's surface, they may fall on top of the lakes, which are mainly composed of methane, ethane, and nitrogen (e.g., \citealt{2016ITGRS..54.5646M,2018E&PSL.496...89M,2020JGRE..12506558P}). If the haze particles float on top of the lakes, they could potentially dampen the surface waves \citep{2019NatGe..12..315C}, which may explain the smooth and almost waveless lakes observed by Cassini \citep{2010GeoRL..37.7104S,2011Icar..211..722B,2012Icar..220..744S,2014GeoRL..41..308Z,2017E&PSL.474...20G}. If they sink, they will sediment to the bottom of the lake and become a part of the lakebed sediment. Based on the work of \citet{2019NatGe..12..315C}, \citet{2020ApJ...905...88Y} determined the flotation criteria for aerosol-sized particles on Titan's lakes: 1) if the aerosol particles are less dense compared to the lake liquids, they will float; 2) if the aerosol particles are denser compared to the lake liquids, they will float if the contact angle formed between the aerosols and the lake liquids is above 0$^\circ$; 3) otherwise, they will sink into the lakes. So far, among all the tholin samples measured here, only the density of the JHU-Plasma sample has been determined, which is around 1400 kg/m$^3$. However, the density of some tholins has been measured to be as low as 500 kg/m$^3$ \citep{2013ApJ...770L..10H}. The density of the lake liquids is around 450--700 kg/m$^3$ \citep{2019NatGe..12..315C}. If the haze particles on Titan are on the lower-end of the density spectrum, they could float on Titan's lakes based on criteria 1. Here, we assume that Titan haze particles are on the high density end of the tholin samples and are denser than the lake liquids, and we use criteria 2 and 3 to determine the floatability of haze particles on Titan's lakes. 

Using the derived surface energies of the tholin samples in Table \ref{table:SE} and the surface tension values for each individual lake component in Table \ref{table:hydrocarbons}, the theoretical contact angles formed with the lake liquids and all tholin samples studied here can be calculated using Equation \ref{eq:thetacalc}. The calculated theoretical contact angles between the lake liquid components and all tholins are shown in Table \ref{table:predicted angles}. Among all the measured tholin samples, we find a 0$^\circ$ contact angle between liquid methane/liquid nitrogen and the tholin samples. For liquid ethane, the UV samples could have non-zero contact angles up to 20$^\circ$ when considering the standard deviations of their measured surface energies. Tholins with the ``average" and the ``end member" surface energies have zero contact angles with all lake liquid components. Assuming these tholins are a good representation of the haze particles, the haze particles are likely completely wetted by the lake liquids, unless the lakes are very ethane-rich and the haze particles are similar to the UV samples.

The high dispersive component among all the tholin samples can lead to similar behaviors towards methane/ethane cloud formation and aerosol-lake interactions on Titan. Because both methane and ethane are non-polar substances, their polar components of surface energies/surface tensions are zero. Thus, when predicting the interactions between these non-polar substances with tholins through the wetting theory (Equation \ref{eq:thetacalc}), the magnitude of the dispersive components of the tholins is the only determining factor. The small contact angles formed between tholins and hydrocarbons can then be attributed to the shared characteristic all tholin samples studied here have: the high dispersive component, $\gamma^d_{sv}$. Our results imply that, on Titan, \textit{it is very likely that methane and ethane clouds can nucleate effectively with the actual haze particles as the CCN, and haze particles that fall towards the hydrocarbon lakes would likely sink to the bottom, if they are denser than the lake liquids.}

\begin{table}
\caption{Predicted contact angles between all the tholin samples and the solid and liquid hydrocarbon condensates (methane and ethane) and the lake liquid components (liquid nitrogen, methane, and ethane) in $^\circ$. For all the ARC tholin samples the contact angles were calculated to be $\theta=0^\circ$ with all these substances.}

\label{table:predicted angles}
\centering
\begin{tabular}{l cc r ccc}\toprule

 \multirow{3}{*}{Sample} &  \multicolumn{6}{c}{Predicted Contact Angles, $\theta$ ($^\circ$)}\\
 \cmidrule{2-7}
 &  \multicolumn{2}{c}{With Solid Hydrocarbon Condensates} && \multicolumn{3}{c}{With Liquid Hydrocarbon Condensates/} \\
  &  \multicolumn{2}{c}{} && \multicolumn{3}{c}{Lake Liquid Components} \\

 \cmidrule{2-3}  \cmidrule{5-7}

 & Methane (s) & Ethane (s)  &&  Methane (l) & Nitrogen (l) & Ethane (l) \\

\midrule
UNI-UV& 0 &  39.9 $\pm$ 6.8 && 0 & 0 & 0--12.9\\

JHU-UV & 0 & 44.6 $\pm$ 4.0 && 0 & 0  & 0--19.5\\
JHU-Plasma & 0 & 31.3 $\pm$ 4.1 && 0 & 0 & 0\\

ARC-Plasma & 0 & 0 && 0 & 0& 0\\
\midrule
Average surface energy & 0 & 31.7 && 0 & 0 & 0\\
End member surface energy -- low $\gamma_s$ & 0& 44.6 && 0& 0 & 0 \\
End member surface energy -- high $\gamma_s$ &0 & 0 && 0 & 0& 0\\
\bottomrule
\end{tabular}
\end{table}

\section{Conclusion} \label{conclusion}
In this study, we measured the surface energy, an intrinsic material property, for a range of Titan haze analogs, or ``tholins", produced in three different laboratories, and performed a thorough cross-comparison study between the different samples made under different experimental conditions. We found that the seven tholin samples studied all have total surface energies that are relatively high, ranging from 47.0--73.3 mJ/m$^2$, indicating high cohesiveness. Despite all the different experimental parameters and laboratory setups used to produce these tholin samples, we were able to identify a commonality between all the tholin samples: a high dispersive component, which may be a characteristic that is shared by the actual haze particles on Titan. Such a common trait would imply that the haze particles on Titan are easily wettable by non-polar hydrocarbon liquids/solids. The differences between the tholin samples are mainly their polar surface energy components. Among all the different experimental conditions used to produce the tholin samples, we identified one key experimental condition, the energy source, to be likely the most important contributor in determining the surface energy of tholins. 

Using the measured surface energies of the tholin samples, we extracted their commonalities and end member properties and proposed an ``average" and two ``end member" surface energies. Calculations involving these proposed surface energies of tholins could respectively represent the likely and extreme scenarios happening on Titan. Using the average and end member surface energies of tholins, we explored the possible implications to further understand two physical processes involving the haze particles on Titan: cloud formation and haze-lake interactions. We found that materials with both the average and the two end member surface energies of tholins would be good CCN for hydrocarbon clouds to nucleate on. This indicates that the haze particles on Titan are likely good cloud seeds for methane and ethane clouds, even if they are closer to the tholin samples with extreme properties. We also found that materials with the average and end member surface energies of tholins would all be completely wetted by the lake liquids on Titan, so they would likely sink into the hydrocarbon lakes if they are denser than the lake liquids. This study demonstrates the importance of comparison studies between tholin samples produced in different laboratories, and the need for more laboratory studies, in order to generate a database of material properties of laboratory analogs of Titan's aerosols. Such a laboratory database can assist with analyzing and interpreting the data on the actual haze particles and surface organics that will be returned from future missions such as the rotorcraft Dragonfly mission to Titan \citep{2018LPI....52.2160N}. This study also opens the door for future investigations to better understand cloud formation and haze-lake interactions on Titan.

\section{Appendix A: analytical solutions of the OWRK method}
\label{appendixA}

We can solve Equation \ref{owrk2} analytically for the surface energy of the measuring solid ($\gamma_s$), given the surface tensions ($\gamma_{lv1}$ and $\gamma_{lv2}$) of the two test liquids and their measured contact angles ($\theta_1$ and $\theta_2$) on the solid sample surface:
\begin{equation}
\gamma_{sv}^d = 
\frac{(\gamma_{lv1}^{tot})^2(1+cos\theta_1)^2\gamma_{lv2}^{p} + (\gamma_{lv2}^{tot})^2(1+cos\theta_2)^2\gamma_{lv1}^p - 2\gamma_{lv1}^{tot}\gamma_{lv2}^{tot}(1+cos\theta_1)(1+cos\theta_2)\sqrt{\gamma_{lv1}^p\gamma_{lv2}^p}}{4(\sqrt{\gamma_{lv1}^d\gamma_{lv2}^p}-\sqrt{\gamma_{lv2}^d\gamma_{lv1}^p})^2}
\end{equation}

\begin{equation}
\gamma_{sv}^p =
\frac{(\gamma_{lv1}^{tot})^2(1+cos\theta_1)^2\gamma_{lv2}^{d} + (\gamma_{lv2}^{tot})^2(1+cos\theta_2)^2\gamma_{lv1}^d - 2\gamma_{lv1}^{tot}\gamma_{lv2}^{tot}(1+cos\theta_1)(1+cos\theta_2)\sqrt{\gamma_{lv1}^d\gamma_{lv2}^d}}{4(\sqrt{\gamma_{lv1}^d\gamma_{lv2}^p}-\sqrt{\gamma_{lv2}^d\gamma_{lv1}^p})^2}
\end{equation}

\begin{equation}
\gamma_{sv}^{tot} = \frac{(\gamma_{lv1}^{tot})^2(1+cos\theta_1)^2\gamma_{lv2}^{tot} + (\gamma_{lv2}^{tot})^2(1+cos\theta_2)^2\gamma_{lv1}^{tot} - 2\gamma_{lv1}^{tot}\gamma_{lv2}^{tot}(1+cos\theta_1)(1+cos\theta_2)(\sqrt{\gamma_{lv1}^p\gamma_{lv2}^p}+\sqrt{\gamma_{lv1}^d\gamma_{lv2}^d})}{4(\sqrt{\gamma_{lv1}^d\gamma_{lv2}^p}-\sqrt{\gamma_{lv2}^d\gamma_{lv1}^p})^2},
\end{equation}
where $\gamma_{l1}^{tot}$ and $\gamma_{l2}^{tot}$ are the total surface tensions of the test liquids, $\gamma_{l1}^{d}$ and $\gamma_{l2}^{d}$ are the partitioning dispersive components of liquid 1 and liquid 2, and $\gamma_{l1}^{p}$ and $\gamma_{l2}^{p}$ are the partitioning polar components of liquid 1 and liquid 2.

The uncertainties of the calculated solid surface energy come from the measurement uncertainty of the contact angles ($\sigma_{\theta_1}$ and $\sigma_{\theta_1}$). Thus, we can use the propagation of uncertainty to analytically derive the errors of the total surface energy and the dispersive and polar components for the solid (referred as $\sigma_{\gamma}$):
\begin{equation}
    \sigma_\gamma^2=(\frac{\partial\gamma}{\partial\theta_1})^2+(\frac{\partial\gamma}{\partial\theta_2})^2.
\end{equation}
The results are listed in the following equations for the dispersive ($\sigma_{\gamma_s^d}$), polar ($\sigma_{\gamma_s^p}$), and total surface energies ($\sigma_{\gamma_s^{tot}}$) of the solid:
\begin{equation}
\begin{split}
    (\sigma_{\gamma_{sv}^d})^2 = \{ \sigma_{\theta_1}^2[(1+cos\theta_1)sin\theta_1(\gamma_{lv1}^{tot})^2\gamma_{lv2}^p-\gamma_{lv1}^{tot}\gamma_{lv2}^{tot}sin\theta_1(1+cos\theta_2)\sqrt{\gamma_{lv1}^p\gamma_{lv2}^p}]^2 \\ + \sigma_{\theta_2}^2[(1+cos\theta_2)sin\theta_2(\gamma_{lv2}^{tot})^2\gamma_{lv1}^p-\gamma_{lv1}^{tot}\gamma_{lv2}^{tot}sin\theta_2(1+cos\theta_1)\sqrt{\gamma_{lv1}^p\gamma_{lv2}^p}]^2 \}/4(\sqrt{\gamma_{lv1}^d\gamma_{lv2}^p}-\sqrt{\gamma_{lv2}^d\gamma_{lv1}^p})^4
\end{split}
\end{equation}

\begin{equation}
\begin{split}
    (\sigma_{\gamma_{sv}^p})^2 = \{ \sigma_{\theta_1}^2[(1+cos\theta_1)sin\theta_1(\gamma_{lv1}^{tot})^2\gamma_{lv2}^d-\gamma_{lv1}^{tot}\gamma_{lv2}^{tot}sin\theta_1(1+cos\theta_2)\sqrt{\gamma_{lv1}^d\gamma_{lv2}^d}]^2 \\ + \sigma_{\theta_2}^2[(1+cos\theta_2)sin\theta_2(\gamma_{lv2}^{tot})^2\gamma_{lv1}^d-\gamma_{lv1}^{tot}\gamma_{lv2}^{tot}sin\theta_2(1+cos\theta_1)\sqrt{\gamma_{lv1}^d\gamma_{lv2}^d}]^2 \}/4(\sqrt{\gamma_{lv1}^d\gamma_{lv2}^p}-\sqrt{\gamma_{lv2}^d\gamma_{lv1}^p})^4
\end{split}
\end{equation}

\begin{equation}
\begin{split}
    (\sigma_{\gamma_{sv}^{tot}})^2 = \{
    \sigma_{\theta_1}^2[(1+cos\theta_1)sin\theta_1(\gamma_{lv1}^{tot})^2\gamma_{lv2}^{tot} - \gamma_{lv1}^{tot}\gamma_{lv2}^{tot}sin\theta_1(1+cos\theta_2)(\sqrt{\gamma_{lv1}^d\gamma_{lv2}^d}+\sqrt{\gamma_{lv1}^p\gamma_{lv2}^p})]^2 \\ + \sigma_{\theta_2}^2[(1+cos\theta_2)sin\theta_2(\gamma_{lv2}^{tot})^2\gamma_{lv1}^{tot} - \gamma_{lv1}^{tot}\gamma_{lv2}^{tot}sin\theta_2(1+cos\theta_1)(\sqrt{\gamma_{lv1}^d\gamma_{lv2}^d}+\sqrt{\gamma_{lv1}^p\gamma_{lv2}^p})]^2
\}/4(\sqrt{\gamma_{lv1}^d\gamma_{lv2}^p}-\sqrt{\gamma_{lv2}^d\gamma_{lv1}^p})^4
\end{split}
\end{equation}

The standard deviation of the calculated $\theta$ in Equation \ref{eq:thetacalc} can be calculated through propagation of error (given that all $\gamma_l^p=0$) when $\theta\neq0$:
\begin{equation}
    \sigma_\theta = \frac{\sigma_{\gamma_{sv}^d}\sqrt{4\gamma_{lv}^d/(\gamma_{lv}^{tot})^2}} {2\sqrt{\gamma_{sv}^d}\sqrt{1-(\sqrt{4\gamma_{lv}^d\gamma_{sv}^d/(\gamma_{lv}^{tot})^2}-1)^2}},
\label{eq:theta_dev}
\end{equation}

\section{Appendix B: extended contact angle with time for diiodomethane}

\begin{figure}[ht] 
    \centering
    \includegraphics[width=\textwidth]{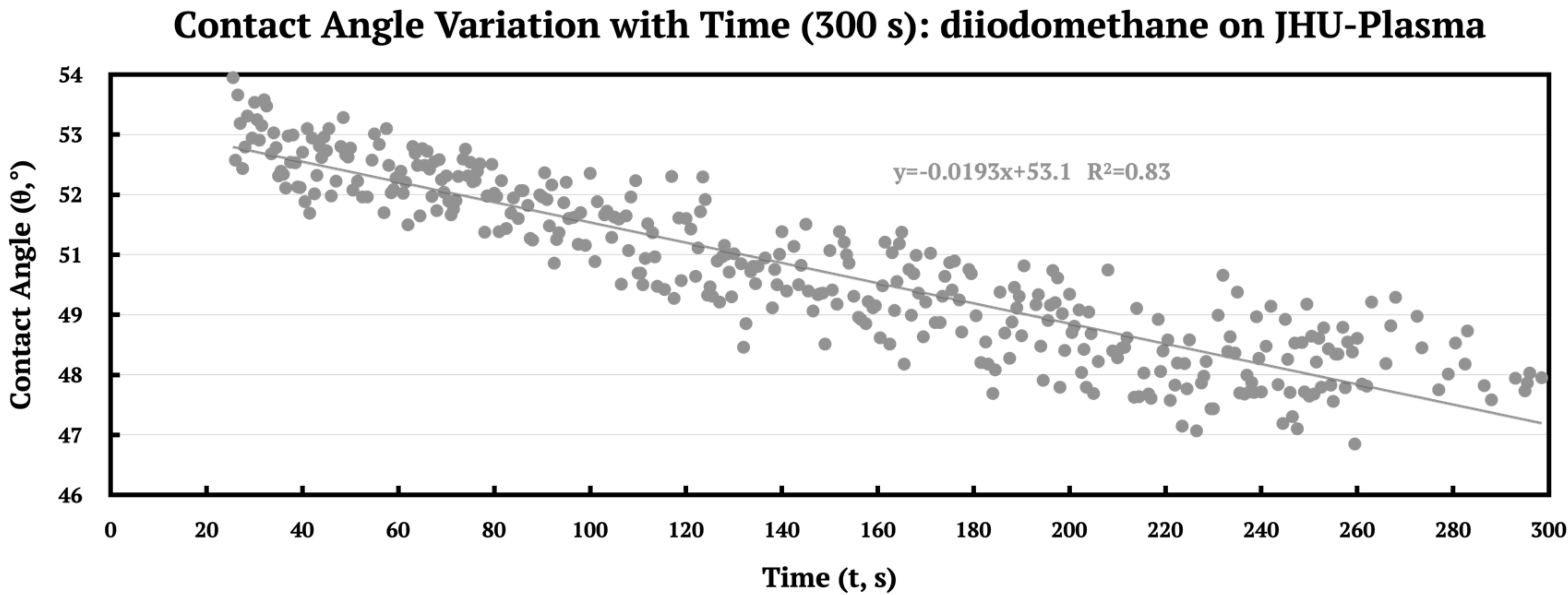}
    \caption{Contact angle variation over 300 s for the JHU-Plasma sample with diiodomethane as the test liquid, along with a linear fit for the decay and the R$^2$ value of the fit.}
    \label{fig:extended}
\end{figure}

Due to the short video recording timescale ($<$60 s) and the slow evaporation of diiodomethane, the decay of contact angle in Figure \ref{fig:60s}b is not readily seen. Thus, we recorded a longer 5 min video and measured the contact angle between diiodomethane and the JHU-plasma sample through time. The results are shown in Figure \ref{fig:extended}. We are able to fit a linear decay of contact angle with time, which is due to solely evaporation of the diiodomethane droplet. The evaporation of diiodomethane is significantly slower than water, the slope of the contact angle decay ($\sim$0.02 $^\circ/s$) is an order of magnitude smaller than the contact angle decay of water ($\sim$0.1--0.5 $^\circ/s$).

\section{data availability}
The measured contact angles recorded with time, for each liquid droplet on each sample, can be found in the supplementary excel file in the Dryad repository at \dataset[10.7291/D13T16]{https://doi.org/10.7291/D13T16}. Please contact the corresponding author directly for accessing the original videos recorded during the contact angle measurements.

\acknowledgments

J. Li thanks the Other Worlds Laboratory at UC Santa Cruz for summer research support. J. Li is also supported by NASA Cassini Data Analysis Program Grant 80NSSC21K0528. X. Yu is supported by the 51 Pegasi b Postdoctoral Fellowship from the Heising-Simons Foundation. X. Zhang is supported by NASA Solar System Workings Grant 80NSSC19K0791. The research conducted at ARC was supported by a NASA's Science Mission Directorate SERA Directed Work Package. F. Salama and E. Sciamma-O'Brien acknowledge the outstanding technical support of R. Walker and E. Quigley. We thank X. Yang for numerous laboratory work support during the pandemic.

\end{document}